\documentclass[reprint,showpacs,superscriptaddress]{revtex4-1}

\pdfoutput=1

\usepackage{amsmath}
\usepackage{amsfonts}
\usepackage{amssymb}
\usepackage{graphicx}
\usepackage{braket}
\usepackage[colorlinks=true, allcolors=blue]{hyperref}

\begin{document}
\title{Microwave generation and frequency comb in a silicon optomechanical cavity with a full phononic bandgap}
\author{Laura Mercad\'e}
\email{laumermo@ntc.upv.es}
\affiliation{Nanophotonics Technology Center, Universitat Polit\`ecnica de Val\`encia, Camino de Vera s/n, 46022 Valencia, Spain}
\author{Leopoldo L. Mart\'in}
\affiliation{Nanophotonics Technology Center, Universitat Polit\`ecnica de Val\`encia, Camino de Vera s/n, 46022 Valencia, Spain}
\affiliation{Departamento de F\'isica, Facultad de Ciencias, Universidad de la Laguna}
\author{Amadeu Griol}
\affiliation{Nanophotonics Technology Center, Universitat Polit\`ecnica de Val\`encia, Camino de Vera s/n, 46022 Valencia, Spain}
\author{Daniel Navarro-Urrios}
\affiliation{MIND-IN2UB, Departament d’Enginyeria Electr\`onica i Biom\`edica, Facultat de F\'isica, Universitat de Barcelona, Mart\'i i Franqu\`es 1, 08028 Barcelona, Spain}
\author{Alejandro Mart\'inez}
\affiliation{Nanophotonics Technology Center, Universitat Polit\`ecnica de Val\`encia, Camino de Vera s/n, 46022 Valencia, Spain}
\date{\today}
\begin{abstract}
Cavity optomechanics has become a powerful tool to manipulate mechanical motion via optical fields. When driving an optomechanical cavity with blue-detuned laser the mechanical motion is amplified, ultimately resulting in phonon lasing. In this work, we show that a silicon optomechanical crystal cavity can be used as an optoelectronic oscillator when driven to the phonon lasing condition. To this end, we use an optomechanical cavity designed to have a breathing-like mechanical mode vibrating at $\Omega_{m}/2\pi=$3.897 GHz in a full phononic bandgap. Our measurements show that the first harmonic displays a phase noise of -100 dBc/Hz at 100 kHz, which is a considerable value for a free running oscillator. Stronger blue-detuned driving leads eventually to the formation of an optomechanical frequency comb, with lines spaced by the mechanical frequency. We also measure the phase noise for higher-order harmonics and show that, unlike in Brillouin oscillators, the noise is increased as corresponding to classical harmonic mixing. Finally, we present real-time measurements of the comb waveform and show that it can be adjusted to a theoretical model recently presented. Our results suggest that silicon optomechanical cavities could be relevant elements in microwave photonics and optical RF processing, in particular in disciplines requiring low-weight, compactness and fiber interconnection. 
\end{abstract}

\maketitle

\section{Introduction}

\begin{figure*}[htbp]
\center
\includegraphics[width=\textwidth]{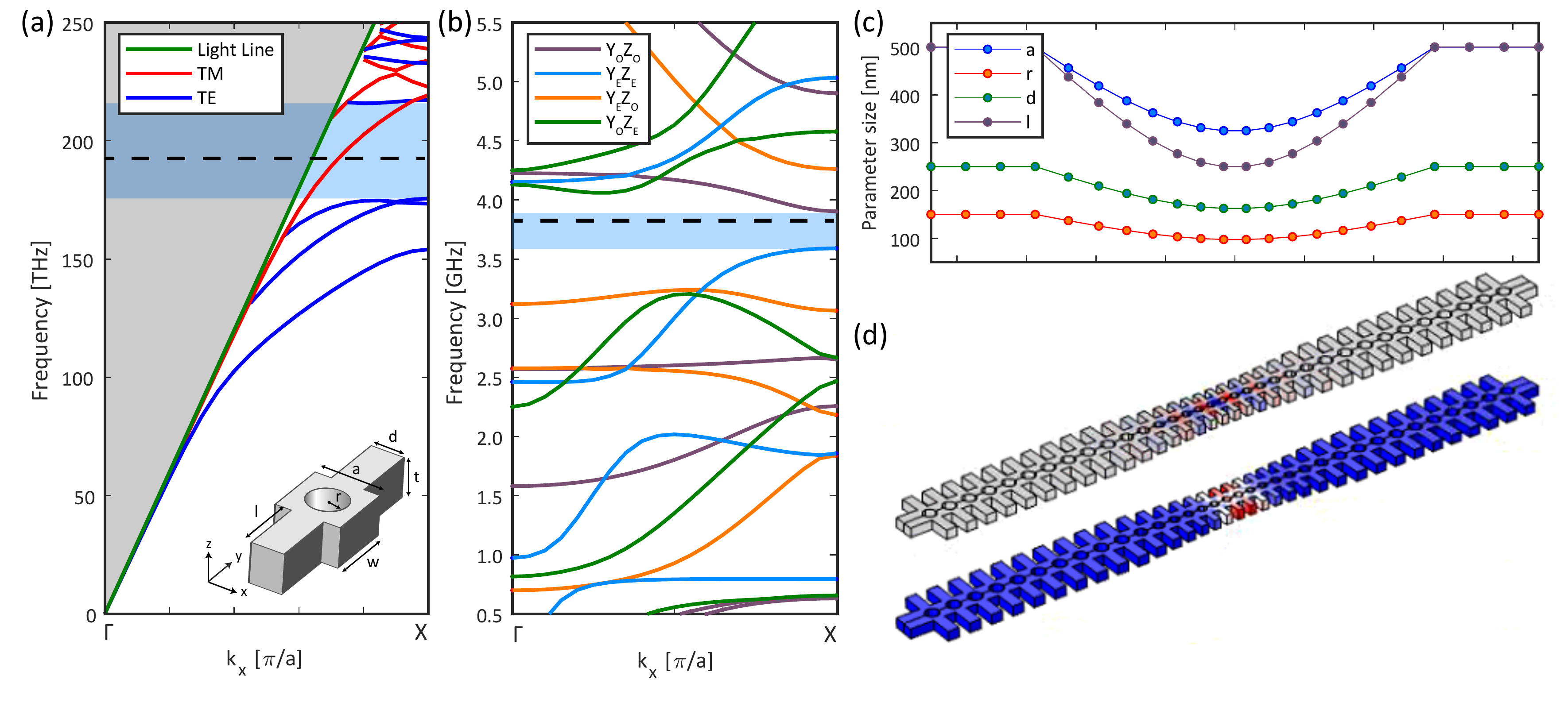}
\caption{Design of a 1DOM cavity with a full phononic bandgap (a) Photonic bands for TE-like modes and TM-likes modes. The grey shaded area denotes the non-guided modes and the blue one the TE-like quasi-bandgap. (b) Phononic bands for the different symmetry mechanical modes. (c) Parameters variation for the nominal structure. (d) Theoretical confined optical mode around $\lambda_{r}=$ 1530 nm and  confined mechanical mode at $\Omega_{m}/2\pi=$3.82 GHz (dashed black line in (b)). The OM coupling rate is $g_{0}/2\pi\simeq$ 540 kHz). The nominal parameters are waveguide w=570 nm, stub mirror (defect) height l=500 nm (250 nm), stub mirror(defect) width d=250 nm (163 nm), hole mirror (defect) radius r=150 nm (98 nm) and mirror (defect) period a=500 nm (325 nm). }
\label{fig:designnominal}
\end{figure*}

Cavity optomechanics addresses the interaction taking place between light and mechanical waves confined in a cavity \cite{KIP08-SCI,ASP14-RMP}. When properly controlled, this interaction can give rise to a plethora of intriguing phenomena such as quantum ground-state cooling \cite{CHAN11-NAT}, phonon lasing \cite{GRU09-PRL,GRU10-PRL}, optomechanically induced transparency \cite{WEIS10-SCI,SAF11-NAT} or non-reciprocal behavior \cite{RUE16-NCOMM}. Amongst the different technological platforms implementing optomechanical cavities (OMCs), their realization in planar photonic integrated circuits enables to accurately design the optical and mechanical resonances as well as maximize its interaction strength (given by the optomechanical (OM) coupling rate, $g_{0}$). This has led to OMCs lithographically defined on released high-index nanobeams -- the so-called OM crystals \cite{EIC09-NAT} -- with mechanical frequencies up to several GHz.

Since the mechanical vibration can efficiently modulate the intensity of an input optical signal, an OMC could be also a key building block in microwave photonics, a discipline that addresses the processing of microwave signals in the optical domain\cite{CAP07-NPHOT}. In this sense, some experiments have shown the capability of OMCs as RF down-converters\cite{HOS08-IEEEPTL} or optoelectronic oscillators\cite{GHOR19-ARX}, which are essential functionalities within microwave photonics. Moreover, since OMCs are nonlinear elements, it has been shown that multiple harmonics of the fundamental mechanical vibrations can be over imposed on the optical signal\cite{KIP05-PRL,CAR05-PRL,HOS08-IEEEPTL}, a phenomenon that has been recently interpreted theoretically as an optomechanical frequency comb (OFC)\cite{MIR18-NJP}. Notably, recent experiments have also shown that frequency combs generated by Kerr nonlinearities could also play a role in the processing of microwave signals in the optical domain\cite{TOR14-LPR}.

In this work, we first demonstrate a OM cavity on a silicon nanobeam having a breathing-mechanical mode vibrating close to 4 GHz with a high $g_{0}$ and placed in a full phononic bandgap. Then, by using blue-detuned laser we demonstrate phonon lasing of this fundamental mechanical mode. We measure the phase noise of the generated microwave tone and show that the OMC can be used as an optoelectronic oscillator. Stronger pumping of the cavity leads to the observation of a series of harmonics forming an OFC in both the optical and RF spectrum, whose phase noise degrades with the harmonic number as in standard harmonic mixing. We also perform real-time measurements of the temporal traces and show that they are in good agreement with a theoretical model of an OFC. This confirms that OMCs can be used for the generation of OFCs and can find application as ultracompact and ultraweight elements for microwave photonics.   

\section{A 1D OM cavity with a full phononic bandgap}
\label{sec:phononicbandgap}

OM cavities can be created on suspended silicon nanobeams with one-dimensional (1D) periodicity. The key idea is to have OM mirrors (that prevent leakage of both photons and phonons) on each nanobeam side whilst at its centre the main parameters of the periodic structure (such as the period or the size of the holes) are adiabatically changed to allocate confined modes. Moreover, the cavity has to be designed to ensure a good overlap between the optical and acoustic resonant modes in order to produce a sufficiently large OM coupling rate $g_{0}$. 

One of the most popular 1D OM cavities in silicon nanobeams \cite{CHAN12-APL} consists of a series of elliptic holes whose size and axial ratio is adiabatically modulated. This OM cavity shows a large value of $g_{0}/2\pi= $ 860 kHz but it does not have a full phononic bandgap but a partial one, because a full phononic bandgap cannot be obtained by merely drilling holes in the nanobeam \cite{PEN11-AIPA}. In order to reduce phonon leakage, the cavity can be surrounded by a two-dimensional (2D) “acoustic shield”, which is basically a 2D structure exhibiting a complete phononic bandgap at the required frequency \cite{CHAN12-APL}. This hybrid 1D/2D approach has been successfully applied in a number of experiments, such as the cooling down to the quantum ground state \cite{CHAN11-NAT,QIU19-ARX}, or phonon guidance through waveguides \cite{FA16-NP},and, more recently, to demonstrate mechanical quality factors $Q_{m}\approx 10^{10} $ in cryogenic environments \cite{MAC19-ARX}. 

Having a full phononic bandgap in a 1D silicon nanobeam requires making lateral corrugations in addition to the holes \cite{PEN11-AIPA}. Using this approach, the existence of mechanical modes in a full phononic bandgap of a 1D OM crystal consisting of circular holes and lateral wings was demonstrated \cite{GOMIS14-NCOMM}. However, the modes within the bandgap in the cavity demonstrated in \cite{GOMIS14-NCOMM} exhibited low $g_{0}/2\pi$ values, being the breathing mechanical mode located out of the phononic bandgap. Thus, the excitation of these modes was not efficient and, in general, all injected energy was used to excite the nanobeam flexural modes (oscillating at tens of MHz), which has been successfully used to demonstrate phonon lasing \cite{NAV15-SR}, chaotic dynamics \cite{NAV17-NCOMM} and, more recently, synchronization \cite{COL19-PRL}. This basic structure can be further engineered so that breathing-like mechanical modes (with $g_{0}/2\pi\simeq$ 540 kHz) appear within the full bandgap, as shown in Ref. \cite{OUD14-PRB}. Figure \ref{fig:designnominal} shows the numerical calculations performed using COMSOL Multiphysics to design the cavity. Figure \ref{fig:designnominal}(a,b) shows the photonic and phononic band diagram for the mirror unit cell used to built the structure where a dashed line has been used to mark out the confined optical and mechanical mode from the defect cell. The inset in Fig. \ref{fig:designnominal}(a) shows the unit cell parameters used in our structure. It can be seen that the structure can be designed to have a mirror that give rise to a TE-like photonic bandgap (Fig. \ref{fig:designnominal}(a)) and a full phononic bandgap (Fig. \ref{fig:designnominal}(b)). The different band colours in Fig. \ref{fig:designnominal}(b) represent the different symmetry (even (E) or odd (O)) bands as a function of the Y or Z symmetry plane. Figure \ref{fig:designnominal}(c) represents the lattice parameters of the tapered cavity. Confined optical \ref{fig:designnominal}(d) and mechanical \ref{fig:designnominal}(d) modes are obtained when modulating the mirror parameters towards the cavity centre. The theoretical mechanical mode is found at $\Omega_{m}/2\pi=$3.82 GHz and the optical mode around $\lambda_{r}$= 1530 nm, with a theoretical optical quality factor $Q_{o}=3\times10^{3}$. In particular, the frequency of the mechanical breathing mode is inside a full phononic bandgap (dashed black line in Fig. \ref{fig:designnominal}(b)).

\begin{figure}[htbp]
\centering
\includegraphics[width=\linewidth]{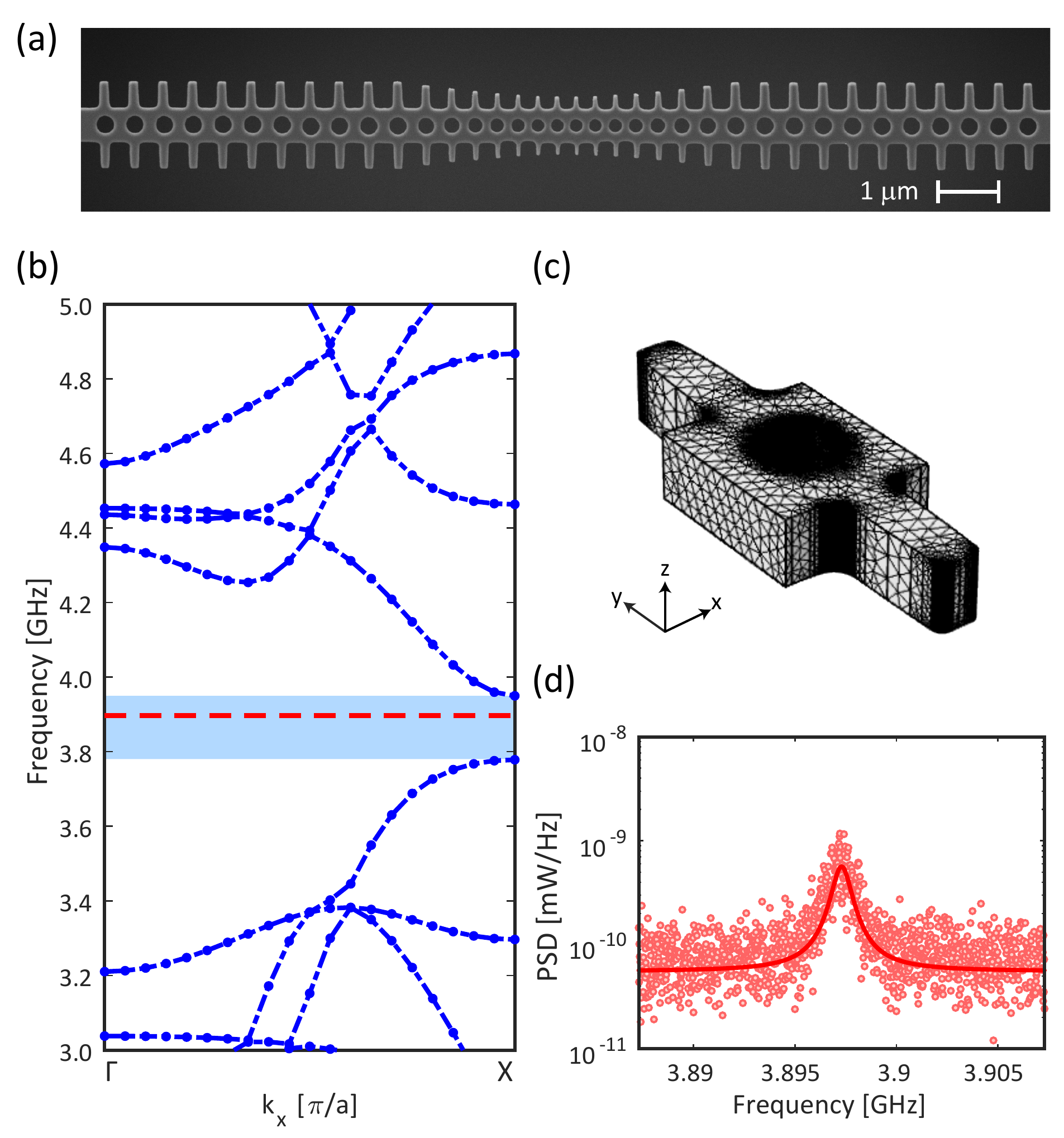}
\caption{Numerical simulations of the fabricated OM cavity (a) SEM image of a fabricated OM cavity which experimentally showed a high-Q mechanical mode. (b) Phononic bands for the real profile mirror unit cell extracted from the SEM image, including the frequency of the experimentally observed mechanical mode, which is well confined inside the phononic bandgap. (c) Unit cell (UC) considered in the phononic bands simulations in COMSOL (d) Power spectral density of the mechanical resonance confined in the total bandgap.}
\label{fig:semstructure}
\end{figure}

A set of OMCs were fabricated following the design summarized in Fig. \ref{fig:designnominal}. Because of fabrication imperfections, the fabricated structures tend were a bit different from the nominal ones. Regarding the optical properties, the fabricated structures had an optical mode at $\lambda_{r}$=(1522.5$\pm$0.3) nm with an optical quality factor of $Q_{o}=5\times10^{3}$ and an overall decay rate of $\kappa/2\pi=39 GHz$. Concerning the mechanical properties, the fabrication imperfections may lead to structures in which the mechanical breathing mode is no longer confined within the mechanical bandgap. To determine whether the fabricated structures also displayed a full phononic bandgap, we simulated numerically an OM cavity using the exact dimensions retrieved from scanning electron microscope (SEM) images, as that shown in Fig. \ref{fig:semstructure}(a). This was done by using a software that enabled us to convert the SEM image into a COMSOL schematic with a unit cell (UC) as that shown in Fig. \ref{fig:semstructure}(c). We simulated different UCs in the mirror region of the cavity, and the final phononic band diagram is presented in Fig.\ref{fig:semstructure}(b). It can be seen that the mechanical bandgap remains even after some fabrication imperfections. In particular, for the frequency $\Omega_{m}/2\pi=$3.897 GHz, which corresponds to the experimentally transduced mechanical mode of the cavity shown in Fig. \ref{fig:semstructure}(d), is within this bandgap. Therefore, we can state that this OM cavity is the first one to show a breathing mode within a mechanical bandgap and with a measured $g_{0}/2\pi=(660\pm 70)$ kHz (See Supplementary Material for more details). In addition, it should be pointed out that in all our experiments we are in the un-resolved sideband regime ($\kappa>\!\!>\Omega_{m}$).

\begin{figure}[htbp]
\centering
\includegraphics[width=\linewidth]{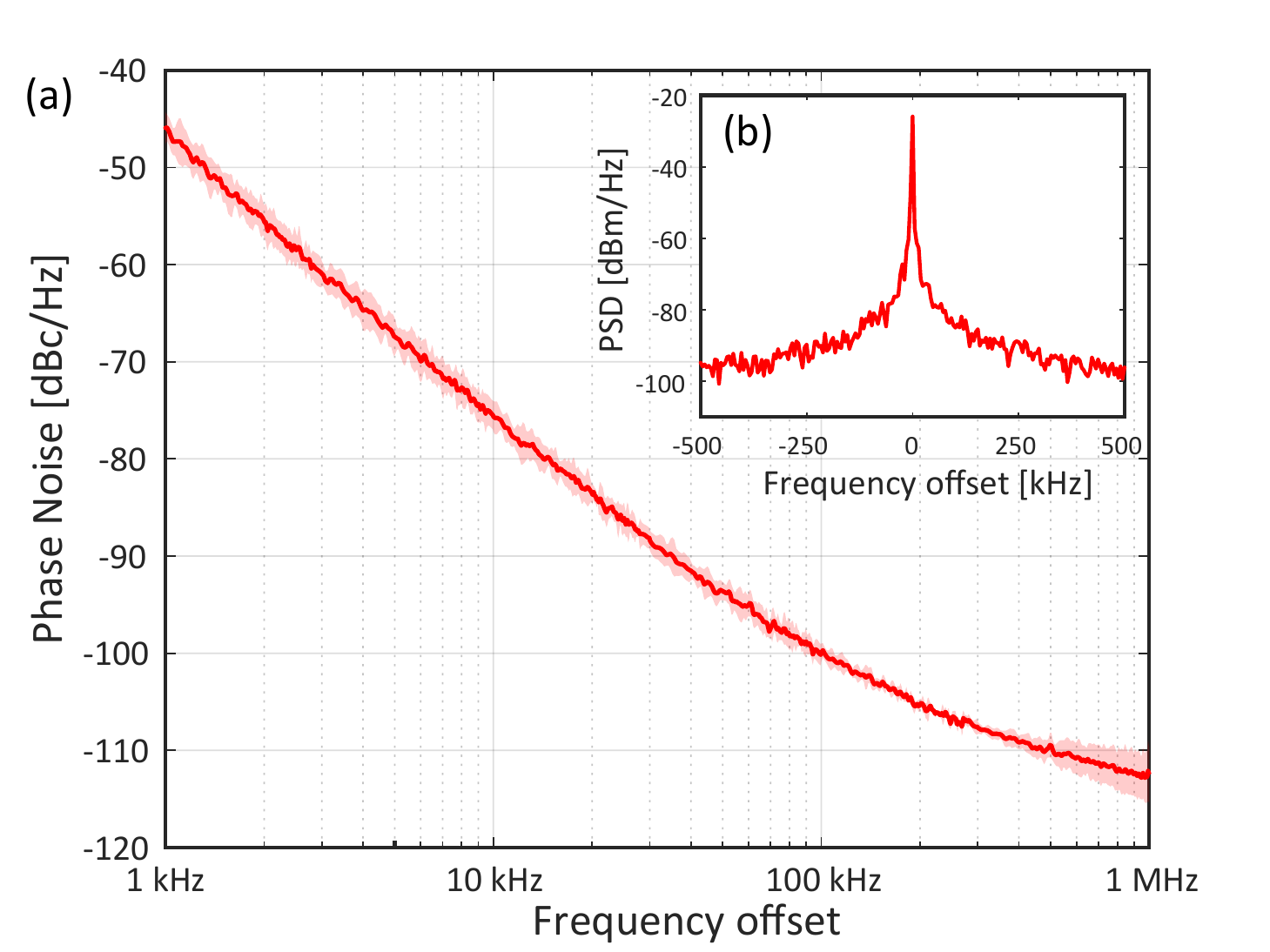}
\caption{OM cavity as an OEO. Generation of a pure RF tone around $\Omega_{m}/2\pi\simeq$ 3.87 GHz (a) Mean phase noise of the generated microwave tones under three different measurements in red and its standard deviation represented by the shadowed area. The noise figure at 100 kHz is (-100$\pm$1) dBc/Hz for the first harmonic. (b) Log-scale of the first harmonic centered at $\Omega_{m}/2\pi\simeq$ 3.87 GHz. The laser is blue-shifted with respect to the optical resonance in order to achieve phonon amplification and, ultimately, phonon lasing, which enables to get very narrow microwave tone. Under this conditions we get an effective mechanical linewidth $\Gamma_{eff}=(17.1\pm 0.8)$ kHz and a quality factor of $Q_{m,eff}=(1.42\pm 0.07)\cdot 10^{6}$   }
\label{fig:phasenoise}
\end{figure}

\section{Optoelectronic oscillator}

\begin{table*}[htbp]
\centering
\caption{Comparison of the phase noise of different OM-based OEOs.}
\begin{tabular}{ccccc}
\hline
Reference & Structure & $f_{RF}$ (GHz) & Noise figure at 100 kHz (in dBc/Hz) & Foot-print \\
\hline
\cite{SID11-OE} & SiN ring resonator & 0.042 & -120 & $\approx 1 mm^{2}$\\
\cite{LUAN14-SR} & Si 2D photonic crystal cavity & 0.112 & -125 & $< 10 \mu m^{2}$\\
\cite{GHOR19-ARX} & 1D III/V OM photonic-crystal cavity & 2.92 & -87 & $< 10 \mu m^{2}$\\
\cite{LI13-NCOMM} & Silica disk & 21.7 & -110 & $>> mm^{2}$\\
This work & 1D Si OM photonic-crystal cavity & 3.87 & -100 & $< 10 \mu m^{2}$\\
\hline
\end{tabular}
  \label{tab:phasenoisecomp}
\end{table*}

High-quality microwave sources are required for a number of applications. Typically, microwave oscillators are made by applying frequency multiplication to an electronic source. This requires a cascade of frequency-doubling stages, which means that the power of the final signal is greatly reduced. Recently, different techniques to produce microwave tones via optical means have been proposed. The resulting device is called an optoelectronic oscillator (OEO)\cite{MAL11-NP,YAO96-JOSAB}. 

Amongst the different techniques to build an OEO, Stimulated Brillouin Scattering (SBS) has proven itself to be extremely useful since it can provide high-frequency RF signals with extreme purity (or extremely narrow linewidth). For instance, cascaded Brillouin scattering on a high-Q silica wedge cavity enabled the synthesis of 21 GHz microwave tone with a record-low phase noise floor value of -–160 dBc/Hz \cite{LI13-NCOMM}. Owing to the equivalence between Brillouin scattering in a waveguide and OM interaction in a cavity \cite{LAER16-PRA}, we could think that an OEO can be also obtained form an OM cavity when pumped with a blue-detuned laser source. We made this experiment with the cavity demonstrated in the previous section. The results are shown on Fig. \ref{fig:phasenoise}(a,b): a high-Q microwave tone at a frequency around $\Omega_{m}/2\pi=$3.87 GHz is clearly seen in the RF spectrum taken in reflection after photodetection of the signal backscattered by the OM cavity. Is has to be noted that that once the cavity is pumped, the mechanical mode shifts in frequency at the same time that the linewidth becomes narrower \cite{NAV14-AIPA,PAN18-ACS}, this is why the tone where the phase noise is measured has shifted down to 3.87 GHz, once we are sure that we are in the lasing regime (See Supplementary Material for more information). The driving conditions are: $P_{in}$ = 4mW, $\lambda_{L}$ = 1523.8 nm. Evidently, blue-detuned driving is needed in order to generate phonons so that, when mechanical losses are overcome, a very narrow tone emerges \cite{NAV14-AIPA}. We measured the phase noise of the generated microwave tone. The results are depicted on  Fig. \ref{fig:phasenoise}(a) where the average phase noise of the first harmonic is shown within its standard deviation calculated from different experimental realizations.

The noise figure becomes as low as (-100$\pm$1) dBc/Hz at 100 kHz, which is a remarkable good value for an OEO oscillating at GHz frequencies. This performance is on par with some commercial mid-range devices, such as the Agilent N5183AMXG that displays a phase noise of -102 dBc/Hz at 100 kHz offset for a 10 GHz frequency \cite{AGIREF}. A summary of the previous OEOs using OM interaction is shown on Table \ref{tab:phasenoisecomp}. As it can be seen, our device shows the best performance if we consider both a micron scale foot-print and an RF frequency in the GHz regime. Notice that our device is a free-running or open-loop OEO: the device generates an RF tone and there is not any feedback loop to improve the frequency stabilization. Introducing feedback (for instance, via a Peltier heater placed below the chip and controlled by the amplitude of the photodetected signal at a given frequency) could enable a closed-loop oscillator with even better figure of merit. Therefore, with the advantages of extreme compactness and Si-technology compatibility, our approach is a very promising candidate to build OEOs.

\section{OM frequency comb}

\begin{figure*}
\center
\includegraphics[width=\linewidth]{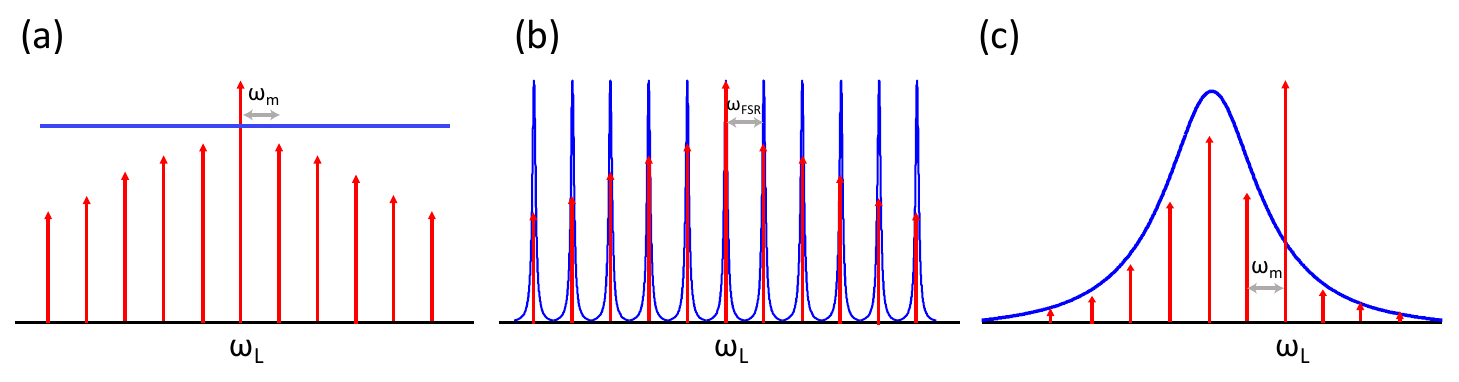}
\caption{Generation of RF harmonics and OFCs in photonic structures. (a) In an optical waveguide, a cascaded Brillouin process gives rise to a set of harmonics spaced $\omega_{m}$ between them; (b) In a travelling-wave resonator supporting a series of high-Q WGMs with a Kerr non-linearity, a set of harmonics is generated via four-wave mixing with a spacing $\omega_{FSR}$ from the laser frequency, where $\omega_{FSR}$ corresponds to the Free spectral Range (FSR) of the resonator ;(c) In an OM cavity, OM interaction induced by a blue-detuned laser scatters photons at frequencies $\omega_{m}$ apart from the laser frequency, which in a cascaded process generated more photons, giving rise to a series of harmonics, and ultimately to an OFCs with a total bandwidth limited by the cavity linewidth. In the figure, the blue curves stand for the photonic response of the structure and the red arrows show the different harmonics.}
\label{fig:harmonicgen}
\end{figure*}

At even higher driving powers, and always operatingwith the laser blue-detuned with respect to the opticalresonance, higher-order harmonics can be observed in the detected signal. The underlying process is similar to cascaded Brillouin scattering in an optical waveguide \cite{KAN09-NP}: the Stokes and anti-Stokes photons (resulting from photon-phonon scattering from the laser source) become new pump signals that generate new Stokes and anti-Stokes photons via photon-phonon interaction, and the process is repeated over and over as long as the resulting photons can propagate in the medium (Fig. \ref{fig:harmonicgen}(a)). This process can be extremely efficient in an optical cavity supporting both optical and mechanical whispery-gallery modes (WGMs) if the mechanical frequency equals the free-spectral range (FSR) \cite{LI13-NCOMM}. In this case, there is a strong resemblance with the formation of optical frequency combs in travelling-wave cavities as a result of third-order nonlinear interaction: the large density of states at the resonant frequencies gives rise to a set of equidistant frequency lines at the output (Fig. \ref{fig:harmonicgen}(b)).

The generation of multiple harmonics in a ``bad'' OM cavity inherits features of both effects (Fig. \ref{fig:harmonicgen}(c)). First, the cavity linewidth is not so broad as in the case of the waveguide, but the high density of states around the resonant frequency benefits the underlying OM interaction. In comparison with the WGM cavity, our cavity is single mode and do not support multiple modes exactly spaced by the mechanical frequency, but this relax the conditions to get multiple harmonics. This way, in the output optical spectrum we expect a series of peaks at frequencies $\omega_{L}+n\omega_{m}$, with amplitudes $A_{m}$, $n$ being an integer number. This closely resembles an optical frequency comb (OFCs) of OM nature, which has been recently analyzed theoretically in \cite{MIR18-NJP}. Notice that, as in the case of OFC based on the Kerr-effecting traveling-wave optical resonators \cite{HAY07-NAT,KIP11-SCI}(Fig. \ref{fig:harmonicgen}(b)), here it is a third-order nonlinear process what mediates in the OFC generation.

Since all generated photons are resonant within the same optical mode (Fig. \ref{fig:harmonicgen}(c)), the number of optical carriers forming the OFC is limited by the linewidth of the optical resonance. As suggested in Ref. \cite{MIR18-NJP}, a low-Q or ``bad'' cavity, meaning that  the cavity works at the un-resolved sidebang regime, should therefore provide wider OFCs than a cavity operating in the sideband-resolved regime. OM-generated OFCs have been previously observed using MHz-scale mechanical modes \cite{NAV17-NCOMM,SAV11-OL,SID11-OE,LUAN14-SR}. In this case, a high-Q cavity with a bandwidth $\approx$1 GHz (Q $> 10^{5}$) can perfectly allocate a large number of harmonics. However, if we want to build an OFC from a GHz-scale fundamental harmonic, the Q factor of the cavity should be reduced to allow the build-up of a sufficient number of harmonics. This is consistent with the theoretical results in Ref. \cite{MIR18-NJP}, in which it is stated that a ``bad'' cavity would perform better than a ``good'' cavity in what refers to OFC generation. In our case, the measured optical Q factor of the cavity was $5\cdot 10^{3}$ at $\lambda_{r}$=(1522.5$\pm$0.3) nm (with an overall decay rate of $\kappa/2\pi=39$ GHz, which satisfies the criteria in Ref. \cite{MIR18-NJP} to produce an OM-OFC with the confined mechanical mode at $\Omega_{m}/2\pi=$3.897 GHz.

\begin{figure}[htbp]
\centering
\includegraphics[width=\linewidth]{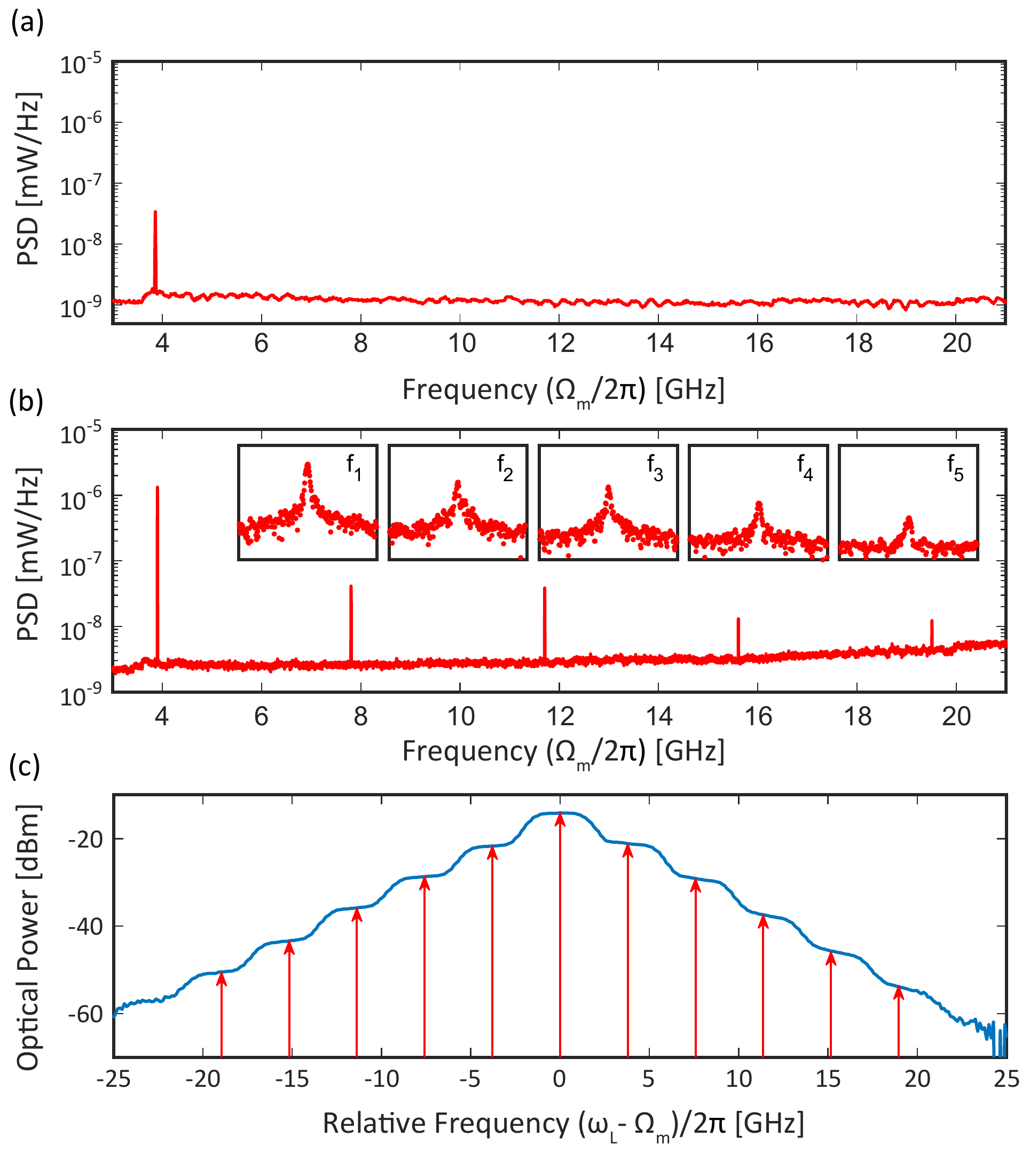}
\caption{OFC generation in the OM cavity. (a) Detected RF spectra at growing driving powers with the laser blue-detuned with respect to the cavity resonance where a single tone is observed at $f_{m}$; (b) RF spectra where the input power and the detuning are increased resulting in the formation of a number of harmonics, reaching the lasing or comb regime. Five harmonics are observed, being higher-harmonic tones obscured by the thermal noise of the detection system. As insets, a close view of the five harmonics within a span of 200 MHz is shown; (c)  Recorded optical spectrum in blue showing a set of peaks corresponding to different harmonics represented in red at the expected position.}
\label{fig:comboptel}
\end{figure}

To perform the experiment, we blue-detuned our laser and recorded the detected RF spectra under different pump powers. As shown in Fig. \ref{fig:comboptel}(a), for lower input powers and highly detuned wavelengths the fundamental frequency corresponding to $f_{m}=\Omega_{m}/2\pi$ is observed. When the driving power is increased, higher-order harmonics are generated, thus resulting in an OM comb, whose first five harmonics are shown in Fig. \ref{fig:comboptel}(b). In figure \ref{fig:comboptel}(c) we also show the optical spectrum of the generated OFC. The width of the peaks observed in the optical spectrum (Fig. \ref{fig:comboptel}(a)) are limited by the resolution of the optical spectrum analyzer, which is 0.01 nm (1.2 GHz). When photodetected, the optical peaks are beaten resulting in the multiple peaks seen in Fig. \ref{fig:comboptel}(b). Notice that any peak results from the addition of many different beats. For instance, the peak at 2$\Omega_{m}$ results from summing up the beating notes $\omega_{L}+2\Omega_{m}$ with $\omega_{L}$, $\omega_{L}-2\Omega_{m}$ with $\omega_{L}$, $\omega_{L}+\Omega_{m}$ with $\omega_{L}-\Omega_{m}$, and all the other optical peaks that are spaced exactly that frequency. However, for any frequency, the terms including a beating of $\omega_{L}$ will dominate.

\begin{figure}[htbp]
\centering
\includegraphics[width=\linewidth]{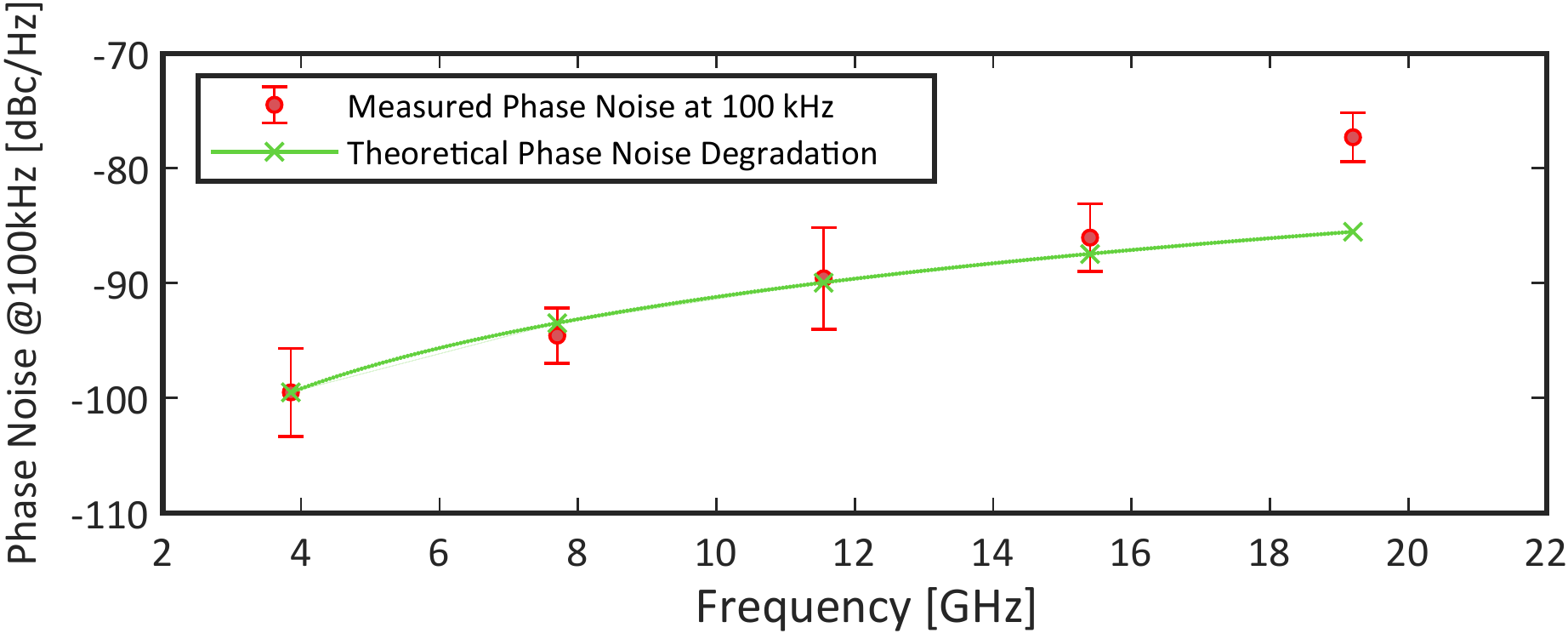}
\caption{Phase noise of the different harmonics of the OFC. In red, the mean value of the phase noise for different phase noise measurements is presented with its standard error and in green, the theoretical phase nose degradation of $20\times log(m)$ with respect to that of the 1st harmonic is shown.}
\label{fig:phasenoisedegradation}
\end{figure}

We have also measured the phase noise of the different harmonics, as shown in Fig. \ref{fig:phasenoisedegradation}. In principle, the harmonic mixing process will result in an added phase noise of $20\times log(m)$ with respect to that of the first harmonic. It can be seen that the previous rule is well satisfied in our device. Notice that this is in stark contrast with the OEO using Brillouin in high-Q cavities, for which higher harmonics show a better noise performance because the Brillouin process ``purifies'' the laser beam \cite{LI13-NCOMM}. This is because the device in Ref. \cite{LI13-NCOMM} operates in the regime where the photon lifetime exceeds the phonon lifetime in the cavity, which results in a narrowing of the Stokes line \cite{SAF19-OPT}. In our case, we operate in the opposite scenario, being optical losses larger than mechanical losses so the scattered Stokes wave is essentially a frequency-shifted copy of the laser beam plus an extra phase noise added by the mechanical oscillator \cite{SAF19-OPT}. Thus, the cavity acts as a harmonic generator (comb) and the phase noise behaves as in a conventional harmonic mixer, being the phase noise impaired by the factor previously mentioned.

In addition, we can also study the optical traces of the generated OM comb in the time domain. The dynamics generation of OM frequency combs can be described by the next set of coupled equations describing the time evolution of the optical mode amplitude $a$ and the amplitude of motion $x$ as \cite{MIR18-NJP,ASP14-RMP}:

\begin{equation}
\dfrac{da}{dt}-(i(\Delta+Gx)-\dfrac{\kappa}{2})a=\sqrt{\kappa_{e}}s_{in}
\label{eq:comb1}
\end{equation}
\begin{equation}
\dfrac{d^{2}x}{dt^{2}}+\Gamma_{m}\dfrac{dx}{dt}+\Omega^{2}_{m}x=\dfrac{\hbar G}{m_{eff}}\vert a\vert^{2}
\label{eq:comb2}
\end{equation}

\noindent where $\Delta$ is the detuning of the laser from the optical resonance, $G$ is the optical frequency shift per displacement, $\kappa$ is the overall intensity decay rate, $\kappa_{e}$ represent the input coupling losses , $s_{in}$ is the input photon flux and $m_{eff}$ the effective mass of the mechanical oscillator.

\begin{figure}[htbp]
\centering
\includegraphics[width=\linewidth]{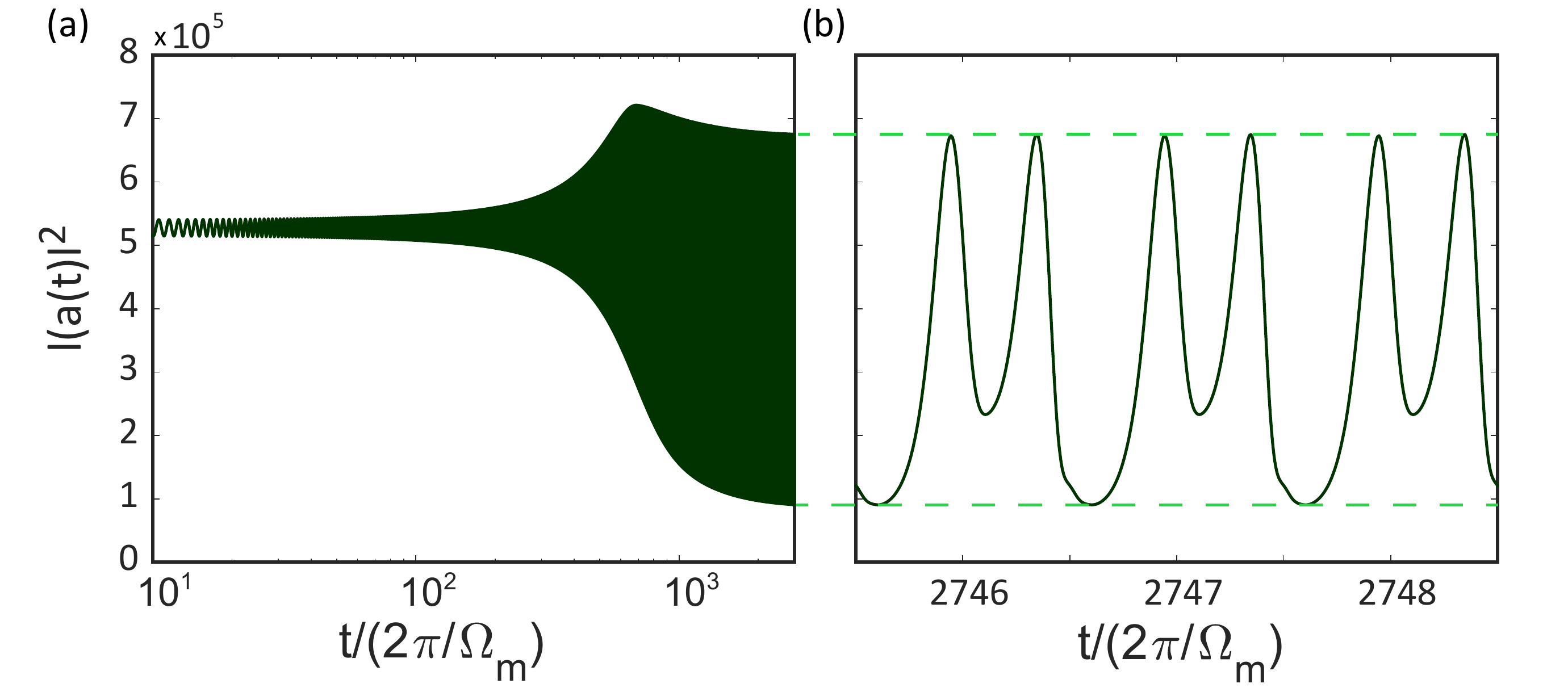}
\caption{(a) Simulated self-induced oscillations showing the intracavity photons evolution versus time in logarithmic scale. (b) Close view of the saturated oscillation with a stable amplitude corresponding to the optical traces of the comb regime.}
\label{fig:oscillations}
\end{figure}

In the blue detuned regime ($\Delta=\omega_{L}-\omega_{r}>$0), as the one we are working with, the optomechanical damping rate $\Gamma_{opt}$ becomes negative, decreasing the overall damping rate $\Gamma_{opt}+\Gamma_{m}$ which firstly leads to heating of the oscillator. In this situation, when the overall damping rate finally becomes negative, an instability appears and an exponential grown of any fluctuation arises, which will finally saturate by nonlinear effects giving rise to a steady-state regime or lasing regime \cite{ASP14-RMP}. The saturation of these self induced oscillations can be seen in Fig. \ref{fig:oscillations}(a) which was obtained from the numerical simulations of the optomechanical coupled equations Eq. \ref{eq:comb1} and Eq. \ref{eq:comb2}. A close view in the saturated regime is shown in Fig. \ref{fig:oscillations}(b).

\begin{figure}[htbp]
\centering
\includegraphics[width=\linewidth]{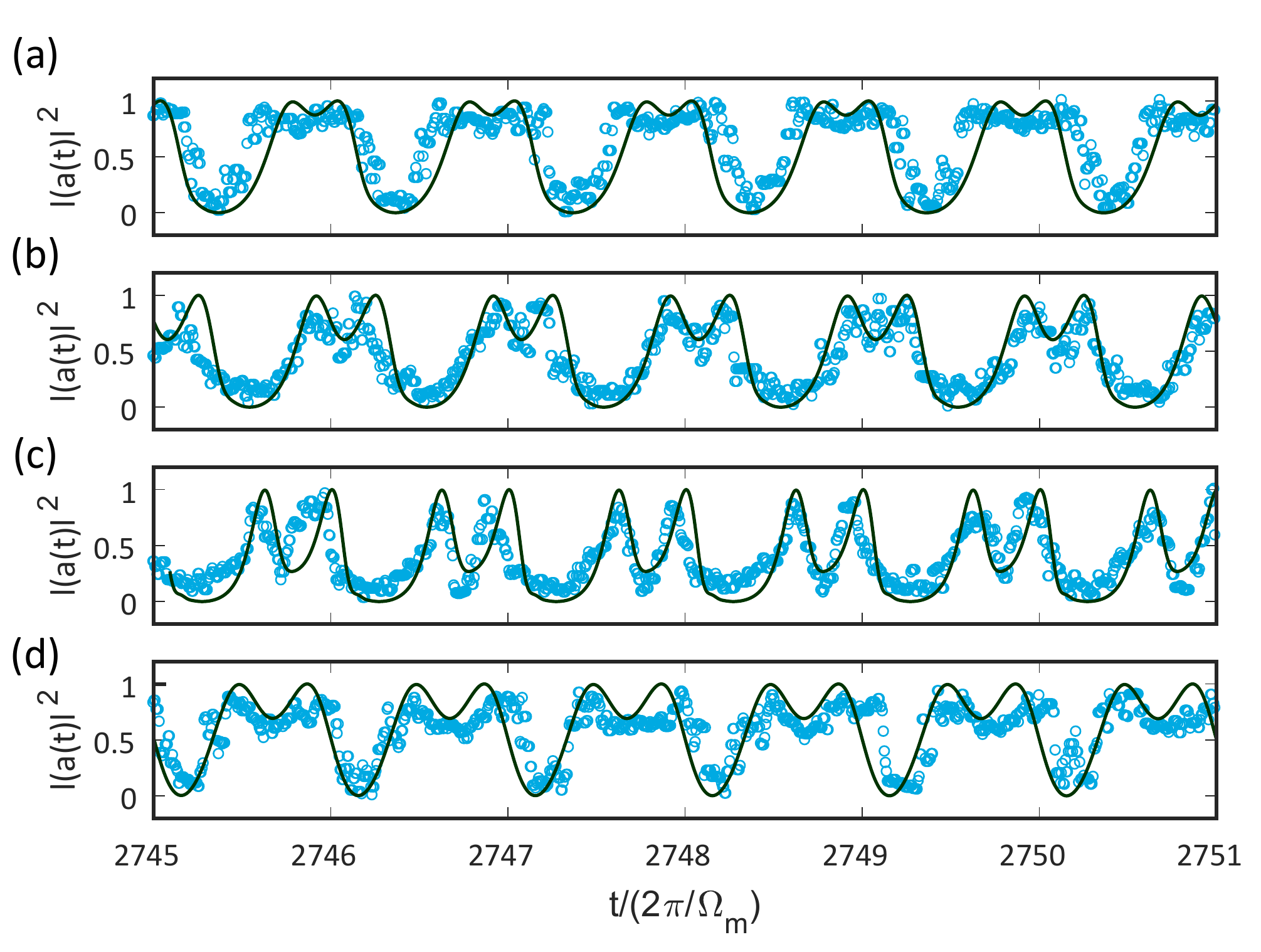}
\caption{Comparison of the experimental (blue dots) and theoretical(solid line) temporal OM-generated OFC traces under different experiment conditions. (a) $\lambda$=1543.35 nm (experiment) and $\Delta=4\Omega_{m}$; (b) $\lambda$=1544.05 nm (experiment) and $\Delta=3.4\Omega_{m}$ (simulation); (c) $\lambda$= 1545.37 nm (experiment) and $\Delta=2.4\Omega_{m}$ (simulation); (d)$\lambda$= 1545.65 nm (experiment) and $\Delta=\Omega_{m}$(simulation).}
\label{fig:opticaltraces}
\end{figure}

The experimental traces were acquired by photodetecting the transmitted optical signal via a high speed and high sensitivity photoreceiver whilst the reflected signal was used to trigger the waveform in a high-speed oscilloscope (for more details about the experimental setup see Supplementary Information). Both, the theoretical and experimental temporal traces are presented in Fig. \ref{fig:opticaltraces} for different laser wavelengths – corresponding to different detunings -- and input photon fluxes $\vert s_{in}\vert^{2}$ –- placed between $0.9\times 10^{17}$ and $1.7\times 10^{17}] s^{-1}$. Despite the wavelength resonance was at 1522.5 nm, we were able to shift it up to values as higher as 1550 nm, as in the recorded traces reported here, due to the thermo-optic effect \cite{NAV14-AIPA}. Besides the experimental parameters, the cavity parameters required in the theoretical model were obtained from simulation of the fabricated structure. In particular, we used a mechanical zero-point fluctuation amplitud $Xzpf=2.69\cdot10^{-15} $m, as well as the value of $m_{eff}=2.7\cdot 10^{-16} $kg, obtained from the OM crystal cavity simulations. For the overall decay rate we introduce the one measured also experimentally $\kappa/2\pi=39 $GHz and $\kappa_{e}=0.26\kappa$.

Figure \ref{fig:opticaltraces} shows a nice agreement between the experiments and the theoretical model. Different waveforms can be obtained by modifying either the detuning or the input power, as pointed out in Ref. \cite{MIR18-NJP}. We observed also different waveforms for other driving conditions, being reproducible using the theoretical model. This confirms that OMCs can be used to synthetize microwave waveforms beyond the generation of single microwave tones and, therefore, they can play a role in the development of microwave photonics applications in silicon photonics technology. 

\section{Conclusions}
In summary, we have demonstrated a new OM cavity implemented on a silicon chip having a large OM coupling rate for a GHz mode within a full phononic bandgap. We have shown that the new OM cavity can perform as an ultracompact OEO at a microwave frequency around 4 GHz. Notice that our device is a free-running or open-loop OEO: the device generates an RF tone and there is no any feedback loop to improve the frequency stabilization. Introducing feedback (for instance, via a Peltier heater placed below the chip and controlled by the amplitude of the photodetected signal at a given frequency) could enable a closed-loop oscillator with even better figure of merit. Operating at cryogenic temperatures would also enormously improve the phase noise \cite{HOS10-IEEEJSTQE} as a result of the enhancement of the mechanical Q factor because of the full phononic bandgap \cite{MAC19-ARX}. Tunability of the resulting microwave signal could be achieved by injection locking to an external optically-modulated tone \cite{HOS08-APL}. In addition, the preliminary demonstration of the OFC paves the way towards synthesis of microwave signals beyond the generation of pure cw tones. The main advantages of the OM cavity approach for RF signal processing is its extreme compactness and low weight, highly desirable in space and satellite applications, and its compatibility with silicon electronics and photonics technology. The resulting optomechanically-generated OFC could also be useful in sensing and spectroscopy application taking profit of the small distance between comb lines which would enable detection using standard electronic equipment. We also envisage that this new OM cavity could be useful in quantum applications since the mechanical Q factor can be extremely increased as a result of the full phononic bandgap.

\section*{Funding Information}

This work was supported by the European Commission (PHENOMEN H2020-EU-713450); Programa de Ayudas de  Investigación y Desarrolo (PAID-01-16) de la Universitat Polit\`ecnica de Val\`encia; Ministerio de Ciencia, Innovación y Universidades (PGC2018-094490-B-C22, PRX18/00126) and 
Generalitat Valenciana (PROMETEO/2019/123, PPC /2018 /002, IDIFEDER/2018/033)

\section*{Acknowledgments}

The authors thank Borja Vidal and Miguel A. Piqueras for technical support. L. M. would like to thank M.-A. Miri for helpful comments.

\noindent\textbf{Disclosures.} The authors declare no conflicts of interest.

\section*{Supplemental Documents}

Supplementary information accompanies this document.

%

\end{document}